\documentclass[preprint,showpacs,preprintnumbers,amsmath,amssymb,prb]{revtex4}
\usepackage{graphicx} % Include figure files
\usepackage{dcolumn}% Align table columns on decimal point
\usepackage{bm}% bold math

\begin{document}
%\preprint{Preprint, April 20, 2004}

\title{Two-Peak Temperature Dependence\\ of the Microwave Surface Impedance\\in Single-Crystalline YBCO Films}
\author{Vladimir Pan}
\email{pan@imp.kiev.ua}
\author{Alexander Kasatkin, Valentin Komashko,\\Constantin Tretiatchenko}
\author{Oleksa Kalenyuk}
\affiliation{Institute for Metal Physics, National Academy of Sciences of Ukraine, 36 Vernadsky Str., Kiev, 03142, Ukraine}

\author{Alexander Ivanyuta}
\author{Gennady Melkov}
\affiliation{Radiophysics Faculty, T. Shevchenko Kiev National University, Kiev, 03127, Ukraine}
\date{\today}

\begin{abstract}
Temperature dependencies of microwave surface impedance, $Z_{s}(T,\omega
)=R_{s}(T,\omega )+iX_{s}(T,\omega)$, were measured for perfect {\it c}-oriented YBCO thin films deposited on CeO$_{2}$--buffered sapphire substrates.
The measurements were performed with a use of three copper cylindrical
resonators operating at $H_{011}$ mode ($f$ = 34, 65, 134 GHz), which
incorporated the studied YBCO films as end plates. The measurements revealed
a distinct two-peak structure of $R_{s}(T)$ and $X_{s}(T)$ dependencies with
peaks at 28--30 K and 50 K. The peaks become smeared at higher frequencies
as well as in applied dc magnetic field ($\sim1$ kOe), while the peak positions remain almost unchanged. For less perfect, e.g., PLD films, $R_{s}(T)$ and $X_{s}(T)$ dependencies are monotonous (power law). The two-peak $Z_{s}(T)$
dependencies for YBCO films differ from those for high quality YBCO single crystals, where only one much broader frequency-dependent peak of $R_{s}(T)$ was detected earlier. The two-peak $Z_{s}(T)$ behavior is believed to be an intrinsic electron property of extremely perfect quasi-single-crystalline YBCO films. A theoretical model is suggested to explain the observed anomalous $Z_{s}(T)$ behavior. The model is based on the Boltzman kinetic equation
for quasiparticles in layered HTS cuprates. It takes into account the
supposed {\it s}+{\it d}--wave symmetry of electron pairing and strong energy dependent relaxation time of quasiparticles, determined mainly by their elastic
scattering on extended defects parallel to the {\it c}--axis (e.g.,
{\it c}--oriented dislocations and twin boundaries).
\end{abstract}

\pacs{74.25.Nf, 74.72.Bk, 74.78.Bz}

\maketitle

\section{INTRODUCTION}

Measurement of the surface impedance $Z_{s}(T,\omega )=R_{s}(T,\omega
)+iX_{s}(T,\omega )$ in high-$T_{c}$ superconductors (HTS) in the microwave
frequency range is one of the most effective and frequently used methods to
study electron properties and mechanisms of superconductivity in these
materials. Such measurements, performed on high quality HTS single crystals or
perfect single-crystalline films, allow to obtain in a
straightforward way the temperature and frequency dependencies of the
complex ac conductivity of the materials $\sigma (T,\omega )=\sigma
_{1}(T,\omega )-i\sigma _{2}(T,\omega )$ in the microwave frequency range,
that in turn yield a complimentary information on microscopic electron
properties of HTS, such as low-energy quasiparticle excitations from the
superfluid condensate, their scattering rate and density of states, the
symmetry of Cooper pairing, etc. Numerous experimental and theoretical
studies of the microwave response carried out during the last decade have
revealed a lot of interesting features of the superconducting state in HTS
metal-oxide compounds and partly shed a light on the nature of
superconductivity in these materials (e.g., {\it d}--wave type of Cooper
pairing).\cite{Bonn'94,Bonn'99,Bonn'01,Bonn'00,Bonn'03,Trunin'98,Hensen'97,tsindl'00,weber'01}

However, up to date there is no comprehensive understanding of microwave
response in HTS. In particular, this concerns the temperature dependence of
surface resistance $R_{s}(T,\omega )$ in highly perfect single crystals
and epitaxial films, where nonmonotonous character of this dependence with a
wide peak below $T_{c}$ was observed by many investigators. Unfortunately, there are still some difficulties in its understanding and explanation in the
framework of existing theoretical models.\cite{hirsch'93,hirsch'00,hirsch'01,hirsch'02,Lee'02,she'03,Ruv'99} 
In the present work we demonstrate for the first time that the nonmonotonous
character of $R_{s}(T)$ in epitaxially-grown single-crystalline YBCO films
can be even more complicated than it was suggested before: in our
experiments $R_{s}(T)$ curves have two distinct rather narrow peaks at quite
different temperatures $T_{1,2}$ ($T_{1}$ = 25--30~K, $T_{2}\approx 50$~K).
This observation clearly indicates that the microscopic scenario of electron
properties in HTS (YBCO) is more intriguing and sophisticated than it was
assumed before.

The temperature dependence of microwave surface resistance, $R_{s}(T)$,
in YBCO perfect single crystals\cite
{Bonn'94,Bonn'99,Bonn'01,Bonn'00,Bonn'03,Trunin'98} and epitaxially grown
single-crystalline thin films\cite{Hensen'97,tsindl'00,weber'01}
observed in a number of experiments performed by different
groups, turned out to be nonmonotonous and revealed a pronounced broad peak at $T\leq T_{c}/2$. The temperature position and height of the peak
depend on frequency and crystal quality. It was shown also 
that the peculiarity of $R_{s}(T)$ is very sensitive to the crystal defect
density. For instance, impurities (point defects) suppress the
peak of $R_{s}(T)$.\cite{Bonn'94,weber'01} Analysis of these
experimental data, based on the phenomenological approach assuming the Drude
form of microwave conductivity for thermally excited quasiparticles $\sigma
(T,\omega )=(n_{q}(T)e^{2}/m)\left[ i\omega +\tau ^{-1}(T)\right] ^{-1}$,\cite{Bonn'99,Bonn'01} sheds a light on the nature of observed 
$\sigma (T,\omega )$ peaks and explains also (at least qualitatively) its
frequency dependence and suppression of the peak by impurities.\cite
{Bonn'94,weber'01} This approach allows also to extract the value of
quasiparticle relaxation time $\tau (T)$ directly from microwave
measurements of $R_{s}(T)$. The $\tau (T)$ value in perfect single crystals
appears to be strongly increasing with the temperature lowering and reaching 
the saturation value of order $10^{-10}$--$10^{-11}$~s at low temperatures (below 20~K).\cite{Bonn'99,Bonn'01}

In the present work the $Z_{s}(T,\omega )=R_{s}(T,\omega )+iX_{s}(T,\omega )$
dependencies are studied experimentally and theoretically for the most
perfect YBCO films in order to establish relation between the microwave
response and the defect nanostructure. The first observation of two-peak
behavior of the $R_{s}(T,\omega )$ as well as $X_{s}(T,\omega )$ dependencies
is presented and a relevant theoretical model is developed.

\section{EXPERIMENTAL}

\begin{figure}
\includegraphics{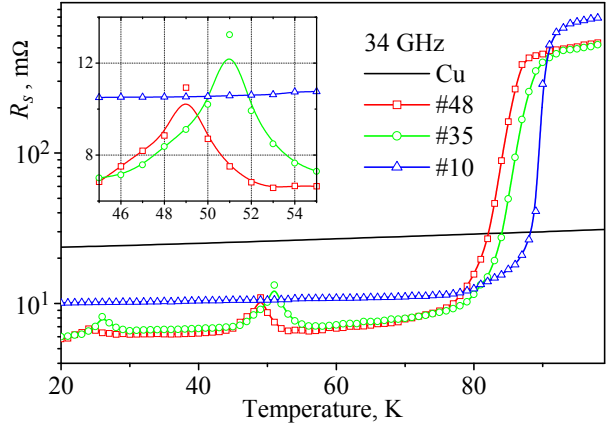}
\caption{\label{fig.1} Temperature dependencies of the surface resistance $R_s(T)$ for three films (\#48, \#35 and \#10) at 34 GHz. $R_s(T)$ for Cu is shown for comparison.}
\end{figure}

\begin{figure}
\includegraphics{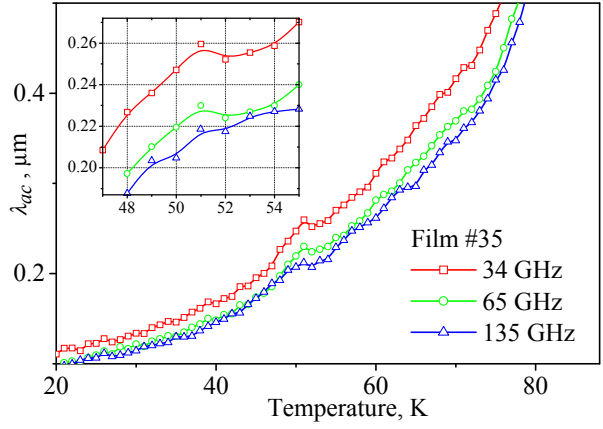}
\caption{\label{fig.4} Temperature dependencies of the penetration depth 
$\lambda _{ac}(T)$ for the film \#35 at three different frequencies.}
\end{figure}

Two-peak temperature dependencies of microwave surface resistance, $R_{s}(T)$, have been observed for the first time in {\it c}--oriented perfect YBCO thin films of various thickness ($d\approx$ 150--480 nm) deposited by off-axis dc magnetron sputtering onto CeO$_{2}$-buffered r-cut single-crystalline
sapphire substrates of 14 $\times$ 14 mm$^{2}$ size. Microwave measurements
were performed with a use of cylindrical pure copper cavities 2, 4 and 8 mm
in diameter. One of flat bases of the cavity was a film under study. The
measurements were performed using $H_{011}$ mode at the frequencies of 134,
65 and 34 GHz. Several dc off-axis magnetron sputtered (MS), as well as
pulse laser deposited (PLD), YBCO films have been measured in the
temperature range from 18 to 100 K. Some films revealed a distinct two-peak
structure of $R_{s}(T)$ and $X_{s}(T)$ dependencies with peaks at 25--30~K
and 48--51~K (Figs.~\ref{fig.1}, \ref{fig.4}). The peaks are much more pronounced at the lower frequency, while their temperature positions remain almost unchanged at any frequency. For less perfect films (e.g., PLD) characterized by a higher density of crystal lattice defects, the $R_{s}(T)$ and $X_{s}(T)$ dependencies appear to be monotonous (power law) and similar to those obtained in previous works.\cite{Hensen'97,weber'01,pan'01,pan'03} The two-peak peculiarity observed for both $R_{s}(T)$ and $X_{s}(T)$ is believed to be an intrinsic electronic feature of perfect quasi-single-crystalline YBCO films. The two-peak behavior is not detected in much smaller YBCO single crystals and in experiments with a use of strip-line resonator measurement technique, which requires film patterning.

\begin{figure}
\includegraphics{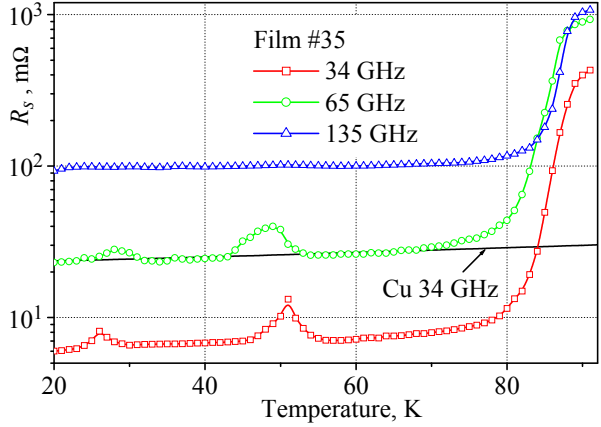}
\caption{\label{fig.2} $R_s(T)$ dependencies for the most perfect film (\#35) at three different frequencies.}
\end{figure}

\begin{figure}
\includegraphics{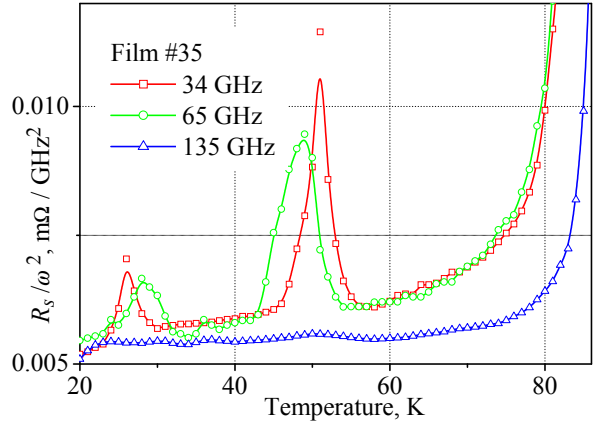}
\caption{\label{fig.3} $R_s(T)$ dependencies for the same film and frequencies as in Fig.~\ref{fig.2} normalized by $\omega ^{2}$.}
\end{figure}

\begin{figure}
\includegraphics{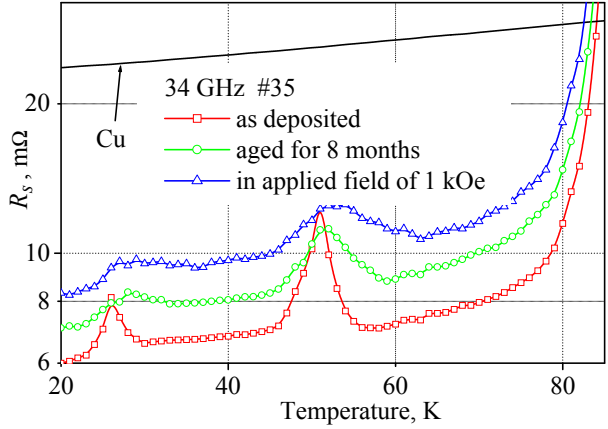}
\caption{\label{fig.5} Effects of aging and applied dc magnetic field on the $R_s(T)$ dependence of the film \#35 at 34 GHz.}
\end{figure}

The observed dependence differs from that for perfect YBCO single crystals, for
which only one much broader frequency dependent peak of $R_{s}(T)$ was
detected.\cite{Bonn'99,Bonn'01,Trunin'98} Some peculiarities of the
two-peak character of $R_{s}(T)$ and $X_{s}(T)$ dependencies are shown in
Figs.~\ref{fig.4}--\ref{fig.5}. The two-peak dependencies of $%
R_{s}(T)$ for one of the most perfect YBCO film are presented in Fig.~\ref{fig.2} for three different frequencies, while Fig.~\ref{fig.3} shows these dependencies normalized by $\omega ^{2}$. The corresponding dependencies of penetration depth $\lambda _{ac}=X_{s}(T)/\omega$ for the same film
are shown in Fig.~\ref{fig.4} for the same three frequencies. Fig.~\ref{fig.5} demonstrates the effect of aging and the influence of applied dc magnetic field on the two-peak structure of $R_{s}(T)$. One can see that the specimen aging as well as application of dc magnetic field lead to smearing of the peaks. Moreover, dc magnetic field shifts the peak positions to slightly higher temperatures. 

It is well known\cite{Bonn'99,Trunin'98} that in superconductors at microwave frequencies and not
too high temperatures ($T~\!<~\!T_c$) usually one has $R_{s}(T,\omega
)\propto\omega ^{2}\sigma _{1}(T,\omega )$, where $\sigma _{1}(T,\omega )$
is the real part of microwave conductivity of superconductor $\sigma
(T,\omega )=\sigma _{1}(T,\omega )-i\sigma _{2}(T,\omega )$. So, the
observed scaling of the $R_{s}(T,\omega )/\omega ^{2}$ curves shown in Fig.~\ref{fig.3} means that $\sigma _{1}(T,\omega )$, which is determined by the
contribution of thermally excited quasiparticles, almost does not
depend on frequency within the studied frequency range: $\sigma _{1}(T,\omega)
\approx {\rm const}(\omega)$ except the vicinity of peaks, where it is falling down rapidly with the increase of frequency. The $\lambda _{ac}(T,\omega )$
behavior is rather similar as it is shown in Fig.~\ref{fig.4}. Thus, our experiments on perfect single-crystalline YBCO films have revealed the new type of nonmonotonous two-peak temperature dependence of $Z_{s}(T,\omega)$, essentially different from the single-peak nonmonotonous behavior observed before for perfect single crystals and films. It should be noted also, that these observed for the first time two peaks of $Z_{s}(T)$ dependence for the most perfect films are much more narrow in the temperature scale than the single peak of $Z_{s}(T)$ for single crystals mentioned above.\cite{Bonn'99,Bonn'01,Trunin'98} Quite similarly to the case of single crystals, these peaks become suppressed, when frequency increases or for less perfect specimens.

The substantial difference between $R_{s}(T,\omega )$ values of YBCO
epitaxially-grown highly biaxially-oriented films and YBCO single crystals
is shown to exist, increasing with temperature, $T\rightarrow T_{c}$. This
difference is supposed to be due to essentially different crystal defect
spectra in YBCO epitaxial films and single crystals.

Perfect quasi-single-crystalline YBa$_2$Cu$_3$O$_{7-\delta}$ films exhibit
several times higher microwave surface resistance than YBCO single crystals.
A different dimensionality of crystal defects in YBCO single crystals and
thin films is supposed to be responsible for the difference. In general, two
major types of crystal defects (point and planar, i.e., oxygen vacancies
and twins) are known to be most essential for electromagnetic behavior of YBCO single crystals. The nonmonotonous $R_{s}(T,\omega)$ dependence with a large broad peak is mostly pronounced in untwined crystals, where only point defects are essential for electron scattering at low temperatures.\cite{Bonn'99,Bonn'01,Bonn'03} In a contrast, {\it c}--oriented extended defects, such as out-of-plane dislocations\cite{Streif'91,chish'92,Merkle'91,pan'97,pan'01,pan'03,Dam'99,Dam'00,Dam'02} and twin boundaries, which in the case of YBCO films usually form much more dense network than in single crystals,\cite{Cas'99,Dur'01,Cao'02,Rob'03} are currently shown to be the most important ones for perfect epitaxially grown YBCO films. Despite the perfect crystallinity, different types of linear defects (dislocations), as well as dislocation arrays, have been identified by TEM/HREM in these films.\cite{Streif'91,chish'92,Merkle'91,pan'97} In order to understand the high-frequency electromagnetic behavior of these films, the existence of edge dislocations arrays should be taken into account, in
particular, out-of-plane edge dislocations, associated with low-angle tilt
dislocation boundaries. TEM/HREM/XRD/AFM characterization of the films under
study revealed a smooth surface (peak-to-valley is 2 nm), high average
in-plane density of out-of-plane edge dislocations ($10^{10}$--$10^{11}$~cm$%
^{-2}$) and a big size of single-crystalline domains ($\langle D\rangle$ =
250~nm), which are separated by low-angle dislocation boundaries. In-plane misalignment of the domains is as low as 0.5--1.0$^{\circ}$.\cite{pan'01,pan'03} 

Basing on this difference of defect structures of YBCO perfect films and single crystals, in the next section we will suggest a model, which can explain (at least qualitatively) the main features of the observed two-peak nonmonotonous behavior of $R_{s}(T)$ dependence in perfect YBCO films, as well as the difference from a single-peak $R_{s}(T)$ dependence in single crystals. 

\section{THEORETICAL MODEL}

The value of microwave surface resistance $R_{s}(T,\omega )$ in a linear regime of microwave response of superconductor at zero applied dc magnetic field is directly determined by the real part $\sigma _{1}(T,\omega )$ of
high-frequency electron conductivity $\sigma (T,\omega )=\sigma
_{1}(T,\omega )-i\sigma _{2}(T,\omega )$ of superconductor: 
\begin{equation}
R_{s}=\frac{1}{2}\,\mu _{0}^{2}\,\lambda _{L}^{3}(T)\,\sigma _{1}(T,\omega ),
\label{Rs}
\end{equation}
while the surface reactance $X_{s}(T,\omega )$ is determined by the ac
penetration depth $\lambda _{ac}(T,\omega )$, which in principle may be
different from the London penetration depth $\lambda _{L}(T)$: 
\begin{equation}
X_{s}(T,\omega )=\mu _{0}\,\omega \lambda _{ac}(T,\omega ),  \label{Xs}
\end{equation}
\begin{equation}
\lambda _{ac}(T,\omega )=\lambda _{L}(T)+\delta \lambda _{n}(T,\omega ),
\label{lambda}
\end{equation}
where $\delta \lambda _{n}(T,\omega )$ is the contribution of excited
quasiparticles to the screening properties of superconductor.

Thus, the observed peaks of $R_{s}(T)$ dependence reveal also the
temperature dependence of $\sigma _{1}(T,\omega )$, because the London
penetration depth $\lambda _{L}(T)$ in Eq.~(\ref{Rs}) is a monotonous function
of temperature. Similarly, peaks in $X_{s}(T)$ and $\lambda _{ac}(T)$
dependencies accordingly to Eqs.~(\ref{Xs}), (\ref{lambda}) are related also to the contribution of the normal component of electron fluid. The $\sigma
_{1}(T,\omega )$ value is generally ascribed to the contribution of 
``normal'' component of electron fluid to the ac conductivity in the
framework of so-called ``two-fluid'' model of superconductor. It should be
mentioned that the phenomenological ``two-fluid'' model, which is
frequently used for description of microwave properties of superconductors, 
\cite{Bonn'94,Bonn'99,Bonn'01,Trunin'98} also follows from the microscopic
BCS theory. A very essential feature of the microscopic approach is that
the role of normal component of electron fluid in superconductor is played by
a gas of Bogolyubov quasiparicles, which are determined as a superposition
of electron and hole states in a normal Fermi liquid. Due to this
circumstance, the normal electron fluid in superconductors has quite
different properties comparatively to those in a normal metal.\cite
{Tinkham'82,Tinkham'96} We will take into account the peculiarities of the
``normal'' electron fluid of Bogolyubov quasiparticles in superconductor and
will show that the principal features of HTS microwave response, including
the observed two peaks of $R_{s}(T)$ and $X_{s}(T)$ dependencies, can be
qualitatively explained using the Boltzman kinetic equation approach for
Bogolyubov quasiparticles with a few additional assumptions about the
symmetry of superconducting state in HTS and its dependence on the concentration
of static defects (impurities, oxygen vacancies, dislocations, etc.).
Namely, we will assume that the case of anisotropic {\it s}+{\it d} pairing is realized\cite{Scalapino'95,Kirtley'00,Yeh'98} (see Fig.~\ref{fig.6}): 
\begin{figure}
\includegraphics{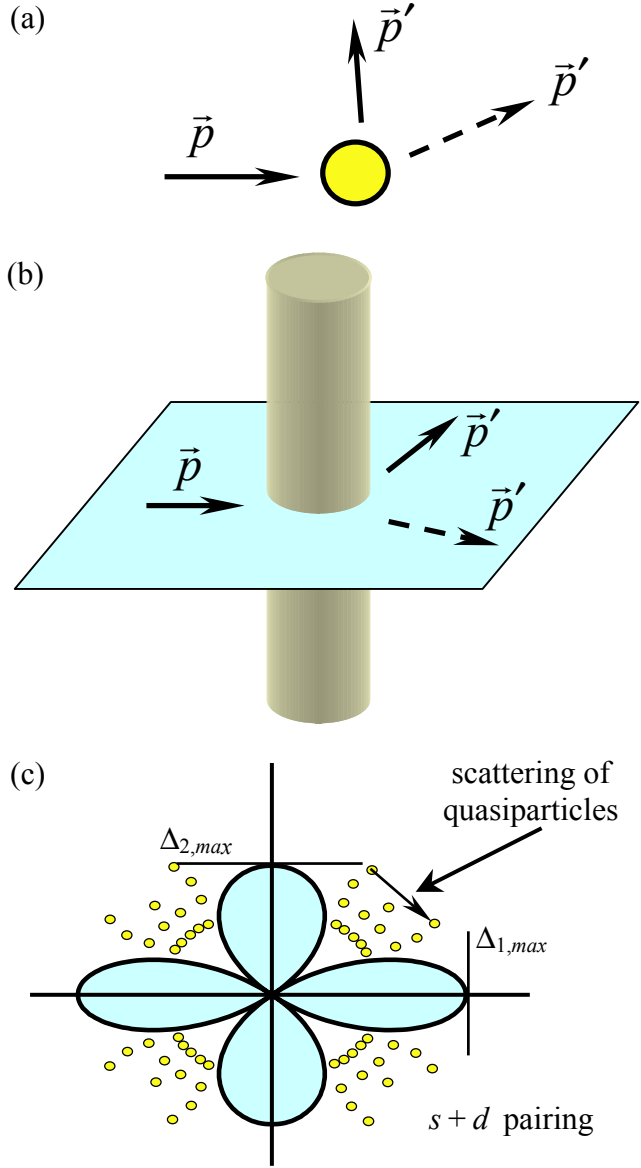}
\caption{\label{fig.6} Schematic representation of anisotropic pairing and scattering processes: (a) 3D scattering on point-like defects; (b) in-plane scattering on extended defects; (c) scattering of quasiparticles in the momentum space in the case of {\it s}+{\it d} pairing.}
\end{figure}
\begin{equation}
\Delta _{{\bf p}}=\Delta _{0}+\Delta _{1}\cos (2\varphi ), \; \varphi =\arctan \left(p_{y}/p_{x}\right), \label{delta}
\end{equation}
and the relation between angle-dependent components $\Delta _{i}$ of the
pairing potential $\Delta _{{\bf p}}$ described by Eq.~(\ref{delta}) is rather sensitive to the defect concentration and to different kind of borders (e.g., film surfaces and twin boundaries).\cite{Kirtley'00,Yeh'98}

In order to calculate the contribution of Bogolyubov quasiparticles to the
conductivity value at microwave frequencies we will use the Boltzman kinetic
equation for nonequilibrium distribution function of quasiparticles $f({\bf %
p)}$. This approach is well known and was widely used for theoretical consideration of electron kinetic properties in normal metals.\cite{Abr} Due to its relative simplicity it allows in principle to take into account  peculiarities of electron spectrum and Fermi surface in HTS\cite{Sch'98,Jar'00} as well as different mechanisms of electron scattering. This approach was argued on the base of microscopic BCS theory in the quasiclassical limit for quasiparticles in superconducting state.\cite{galperin'86,Blatter'99,Kopnin'01}
The Boltzman equation in this case may be written in the form:\cite{galperin'86} 
\begin{equation}
\frac{\partial f}{\partial t}+v_{g}({\bf p})\frac{\partial f}{\partial {\bf r%
}}+e^{*}(\varepsilon _{{\bf p}})E({\bf r},t)\frac{\partial f_{0}}{\partial 
{\bf p}}=-\frac{f-f_{0}}{\tau (\varepsilon _{{\bf p}})},  \label{Boltz}
\end{equation}
where $\varepsilon _{{\bf p}}=\sqrt{\xi _{{\bf p}}^{2}+\Delta _{{\bf p}}^{2}}
$ is the energy of quasiparticles; $\xi _{{\bf p}}={\bf v}_{F}({\bf p}-%
{\bf p}_{F})$ is the energy of a normal electron in a vicinity of the Fermi
surface, ${\bf v}_{F}$ and ${\bf p}_{F}$ are the Fermi velocity and
momentum, respectively, $f_{0}$ is the equilibrium Fermi distribution
function of quasiparticles, $v_{g}({\bf p})={\bf v}_{F}({\bf p)}\,(\xi%
_{\bf p}/\varepsilon _{\bf p})$ is the group velocity of quasiparticles, 
$e^{*}(\varepsilon _{{\bf p}})=e\,(\xi _{{\bf p}}/\varepsilon _{{\bf p}})$
is the quasiparticle charge, which is different from the normal electron charge $e$, because a Bogolyubov quasiparticle is a superposition of electron and hole states in the Fermi liquid,\cite{Tinkham'96} $\tau (\varepsilon _{{\bf p}})$ is the relaxation time of quasiparticles, which in the case of elastic
scattering by static defects can be calculated from the following well known
expression:\cite{galperin'86}
\begin{equation}
\tau ^{-1}(\varepsilon _{{\bf p}})=\frac{2\pi }{\hbar }\int \mid M_{{\bf p,p}%
^{\prime }}\mid ^{2}l^{2}({\bf p},{\bf p}^{\prime })\,\delta (\varepsilon _{%
{\bf p}}-\varepsilon _{{\bf p}^{\prime }})\frac{d^{3}p^{\prime }}{(2\pi
\hbar )^{3}}.  \label{tau}
\end{equation}

Integration in Eq.~(\ref{tau}) is performed in the vicinity of the Fermi
surface. $M_{{\bf p,p}^{\prime }}$ is the matrix element of electron
scattering on the defect. $l({\bf p},{\bf p}^{\prime })$ is the so-called
``coherence factor'': 
\begin{equation}
l^{2}({\bf p},{\bf p}^{\prime })=\left( 1+\frac{\xi _{{\bf p}}\xi _{{\bf p}%
^{\prime }}-\Delta _{{\bf p}}\Delta _{{\bf p}^{\prime }}}{\varepsilon _{{\bf %
p}}\varepsilon _{{\bf p}^{\prime }}}\right),  \label{factor}
\end{equation}
which describes the difference between scattering of quasiparticles
comparatively to usual electrons in the framework of BCS theory\cite
{Tinkham'96} and provides a strong dependence of the relaxation time on the
quasiparticle energy even in the case of elastic scattering on static
defects. In the case of isotropic $s$--wave pairing and scattering on
point-like defects this dependence can be estimated as:\cite{galperin'86}
\begin{equation}
\tau _{i}(\varepsilon _{{\bf p}})\approx \tau _{n} \, 
(\varepsilon _{\bf p}/\xi _{\bf p}).  \label{tau0}
\end{equation}

So, $\tau _{i}(\varepsilon _{{\bf p}})$ diverges at the edge of the gap when 
$p\rightarrow p_{F}$ and the group velocity $v_{g}$ goes to zero. For the
sake of simplicity we have assumed in Eq.~(\ref{tau}) that all defects in the
crystal lattice are identical and can be characterized by the same matrix
element $M_{{\bf p,p}^{\prime }}$. If there are several kinds of scattering
defects, e.g., point-like (oxygen vacancies) and extended ones (dislocations,
twin boundaries) oriented along the $c$--axis, as they usually are in
the case of epitaxial films and as it was discussed in the previous section,
it follows from Eq.~(\ref{tau}) that: 
\begin{equation}
\tau ^{-1}(\varepsilon _{{\bf p}})\approx \tau _{i}^{-1}(\varepsilon _{{\bf p%
}})+\tau _{ext}^{-1}(\varepsilon _{{\bf p}}).  \label{tau1}
\end{equation}

\begin{figure}
\includegraphics{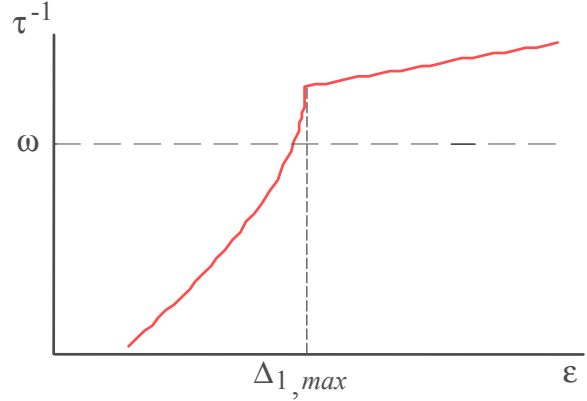}
\caption{\label{fig.7} Energy dependence of the relaxation rate
$\tau ^{-1}(\varepsilon)$ for the case of {\it d}-- or {\it s}+{\it d}--electron pairing and predominant scattering on extended linear defects.}
\end{figure}

The energy dependencies of the relaxation rates in the right hand side of
Eq.~(\ref{tau1}) for point-like and extended defects may be quite different
because point like defects can scatter electrons in all directions of
momentum space, while scattering on extended defects can proceed only with a
conservation of momentum along the $c$--axis. This becomes especially
important for layered HTS materials, where electrons move mainly within
Cu-O layers. In the case of {\it d}--wave or assumed {\it s}+{\it d} (or other type) anisotropic pairing, an additional strong dependence of the relaxation time on the quasiparticles energy $\varepsilon _{{\bf p}}$ has to appear due to
a confinement of momentum space, where the quasiparticle
can be scattered, with a decrease of $\varepsilon _{{\bf p}}$. This effect
follows directly from Eq.~(\ref{tau}) due to the $\delta$--function term in the
integrand. In a certain sense, this effect can be considered as an analogue of Andreev reflection of quasiparticles in the momentum space (Fig.~\ref{fig.6}). The confinement of momentum space for quasiparticle scattering is
mostly essential for the scattering on extended defects, leading to a very
rapid increase of $\tau _{ext}(\varepsilon _{{\bf p}})$, when $\varepsilon _{%
{\bf p}}$ decreases and approaches the value $\Delta _{max}$ ($\Delta
_{max}=\max \left( \Delta _{{\bf p}}\right) $) as it is shown in Fig.~\ref{fig.7}: 
\begin{equation}
\tau _{ext}^{-1}(\varepsilon _{{\bf p}})\propto \arcsin \left( \frac{%
\varepsilon _{{\bf p}}}{\Delta _{max}}\right).  \label{tau2}
\end{equation}
The general solution for the quasiparticle ac conductivity can be obtained
from Eq.~(\ref{Boltz}) by a usual manner.\cite{Ziman'72} The real part 
$\sigma_{1}(T,\omega )$ can be written in form: 
\begin{equation}
\sigma _{1}(T,\omega )=\int \frac{e^{2}v_{g}^{2}(\varepsilon _{{\bf p}})\tau
(\varepsilon _{{\bf p}})}{1+\omega ^{2}\tau ^{2}(\varepsilon _{{\bf p}})}%
\left( -\frac{\partial f_{0}}{\partial \varepsilon }\right) \frac{d^{3}p}{%
(2\pi \hbar )^{3}}.  \label{sigma}
\end{equation}

A strong temperature dependence of $\sigma _{1}(T,\omega )$ as well as
its frequency dependence arise usually due to the 
$\left(-\partial f_{0} / \partial \varepsilon \right)$ term in the integrand of Eq.~(\ref{sigma}). Quite formally this solution can be rewritten in the Drude-like form:
\begin{equation}
\sigma _{1}(T,\omega )=\frac{n_{n}(T) e^{2}}{m}\left\langle \frac{\tau (\varepsilon _{{\bf p}})}{1+\omega
^{2}\tau ^{2}(\varepsilon _{{\bf p}})}\right\rangle _{T},  \label{Drude}
\end{equation}
where $n_{n}(T)$ is the effective concentration of thermally excited
quasiparticles, $\left\langle ...\right\rangle _{T}$ denotes thermal
averaging. As it was originally supposed in Refs.\cite{Bonn'99,Bonn'00,Bonn'03}, the nonmonotonous character and appearance of $R_s(T)$ peak dependence as well as its frequency dependence can be explained properly by a strong increase of quasiparticle relaxation time with temperature lowering according to the Drude expression for the ac conductivity $\sigma _{1}(T,\omega )$. The peak position corresponds to the condition $\tau^{-1}(T)\approx \omega $. We suppose that this explanation is valid also in our case, when two peaks are observed. The emergence of two peaks instead of one can be explained by existence of two different $\Delta _{max}$ values for the assumed case of {\it s}+{\it d} pairing (see Fig.~\ref{fig.6}), while the sharpness of these peaks comparatively to the peak in single crystals is determined by a very strong energy dependence of the relaxation time (Eq.~(\ref{tau2})), when quasiparicles are scattered preferably by extended defects (see Fig.~\ref{fig.7}). This strong energy dependence of $\tau_{ext}$ transforms into sharp peaks of $\sigma_{1}(T)$ according to Eq.~(\ref{Drude}).
   
The present model can explain also some additional features of the microwave
response of YBCO films, such as (i)~smearing of peaks with the frequency
increase, (ii)~lowering of $R_s(T)$ and smearing of peaks with an increase
of point-like defect concentration, (iii)~general quasi-linear behavior of 
$R_s(T)$ at moderate temperatures.

\section{DISCUSSION}

The obtained results shown in Figs.~\ref{fig.1}--\ref{fig.5}, that is the nonmonotonous two-peak structure of $Z_{s}(T,\omega)$ in perfect single-crystalline YBCO epitaxial films, is a strong argument for the scenario of anisotropic electron pairing in HTS. The observed for the first time two-peak peculiarity of $Z_{s}(T)$ dependence for the most perfect single-crystalline YBCO films, as well as the difference from nonmonotonous $Z_{s}(T,\omega)$ dependence for perfect single crystals (and also some less perfect films), can be explained, using just two assumptions: (i)~the anisotropic {\it s}+{\it d} character of electron pairing and (ii)~the dominant role of extended {\it c}--oriented defects in electron scattering processes. These assumptions look quite natural with regard to thin films, where surfaces and/or twin boundaries can lead to more complicated character of electron pairing than the pure {\it d}--wave pairing in perfect single crystals.\cite{Kirtley'00,Yeh'98} On the other hand, the extended $c$--oriented linear or planar defects (most probably, out-of-plane edge dislocations and twins) can play a dominant role in electron scattering. In the case of untwined single crystals there are no extended defects. Therefore, only point defects are essential for electron scattering at low temperatures. The above two assumptions, which seem to be specific for thin films, distinguish them from single crystals and, thus, provide the difference in microwave response: one broad peak of $R_{s}(T)$ for single crystals and two sharp peaks for perfect films. For less perfect films with a higher number of point defects and in the case when only pure {\it d}--wave pairing takes place, their behavior at microwave frequencies is rather similar to that of single crystals.

The observed two peaks of $R_{s}(T)$ are much more narrow comparatively to the single peak for perfect single crystals due to different defect structures in films and single crystals as it was discussed above. Namely, a large number of extended defects along with an anisotropic pairing can lead to emergence of a sharp peak at $T\sim \Delta _{max}(T)$ as it follows from Eq.~(\ref{sigma}). The second peak at a lower temperature is caused by the anisotropy of pairing potential (existence of two different values $\Delta _{max}$ for different directions in momentum space in the case of {\it s}+{\it d} electron pairing as it is shown schematically in Fig.~\ref{fig.6}). The present model for quasiparticle conductivity allows also to understand the frequency dependence of the observed peculiarities and their smearing, when the number of point-like defects increases leading to an increase of
$\tau_{i}^{-1}(\varepsilon _{{\bf p}})$. It should be noted, that in a contrast to suggestions made in some theoretical works, it does not seem to be
necessary to take into account a contribution of inelastic quasiparticle scattering by collective excitations (magnons, phonons, etc.). The two-peak
temperature dependence of $\sigma _{1}(T,\omega )$ can occur in accordance to Eq.~(\ref{Drude}) even in the case of elastic scattering by extended
static defects due to additional effect of anisotropy caused by anisotropic
{\it s}+{\it d}--wave pairing, which leads to the confinement of  momentum space available for scattered quasiparticles as it was discussed above.

\section{CONCLUSION}

The observed two-peak character of $R_s(T)$ dependence is an feature of the most perfect quasi-single-crystalline YBCO films, characterized by a smooth surface, low concentration of defects, large domain size and low-angle
boundaries between them. The $R_s(T)$ dependence in less perfect films is
monotonous. We suppose that the two-peak character of $R_s(T)$, observed
experimentally for the first time, is an intrinsic fundamental property and reveals the peculiarities of anisotropic electron pairing, which manifests in the microwave electron response and can be properly described using the Boltzman
kinetic equation approach for Bogolyubov quasiparticles.


\begin{thebibliography}{99}

\bibitem{Bonn'94} D.~A.~Bonn, S.~Kamal, K.~Zhang, R.~Liang, D.~J.~Baar, E.~Klein, and W.~N.~Hardy, Phys. Rev. B {\bf 50}, 4051 (1994).

\bibitem{Bonn'99} A.~Hosseini, R.~Harris, S.~Kamal, P.~Dosanjh, J.~Preston,
R.~Liang, W.~N.~Hardy, and D.~A.~Bonn, Phys. Rev. B {\bf 60}, 1349 (1999).

\bibitem{Bonn'01} R.~Harris, A.~Hosseini, S.~Kamal, P.~Dosanjh, R.~Liang,
W.~N.~Hardy, and D.~A.~Bonn, Phys. Rev. B {\bf 64}, 064509 (2001).

\bibitem{Bonn'00} A.~J.~Berlinsky, D.~A.~Bonn, R.~Harris, and C.~Kallin, Phys.
Rev. B {\bf 61}, 9088 (2000).

\bibitem{Bonn'03} P.~J.~Turner, R.~Harris, S.~Kamal, M.~E.~Hayden, D.~M.~Broun,
D.~C.~Morgan, A.~Hosseini, P.~Dosanjh, G.~Mullins, J.~S.~Preston, R.~Liang,
D.~A.~Bonn, and W.~N.~Hardy, Phys. Rev. Lett. {\bf 90}, 237005 (2003).

\bibitem{Trunin'98} M.~R.~Trunin, J. Superconductivity {\bf 11}, 381 (1998).

\bibitem{Hensen'97} S.~Hensen, G.~M\"{u}ller, C.~T.~Rieck, and K.~Scharnberg, Phys. Rev. B {\bf 56}, 6237 (1997).

\bibitem{tsindl'00} M.~I.~Tsindlekht, E.~B.~Sonin, M.~A.~Golosovsky, D.~Davidov, X.~Castel, M.~Guilloux-Viry, and A.~Perrin, Phys. Rev. B {\bf 61}, 1596 (2000).

\bibitem{weber'01} J.~Einfeld, P.~Lahl, R.~Kutzner, R.~W\"{o}rdenweber, and
G.~K\"{a}stner, Physica C {\bf 351}, 103 (2001).

\bibitem{hirsch'93} P.~J.~Hirschfeld, W.~O.~Putikka, and D.~J.~Scalapino, Phys.
Rev. Lett. {\bf 71}, 3705 (1993); Phys. Rev. B {\bf 50}, 10250 (1994).

\bibitem{hirsch'00} M.~H.~Hettler and P.~J.~Hirschfeld, Phys. Rev. B {\bf 61},
11313 (2000).

\bibitem{hirsch'01} D.~Duffy, P.~J.~Hirschfeld, and D.~J.~Scalapino, Phys. Rev.
B {\bf 64}, 224522 (2001).

\bibitem{hirsch'02} W.~A.~Atkinson and P.~J.~Hirschfeld, Phys. Rev. Lett. 
{\bf 88}, 187003 (2002).

\bibitem{Lee'02} A.~C.~Durst and P.~A.~Lee, Phys. Rev. B {\bf 65}, 094501
(2002).

\bibitem{she'03} D.~E.~Sheehy, Phys. Rev. B {\bf 68}, 054529 (2003).

\bibitem{Ruv'99} C.~T.~Rieck, K.~Scharnberg, and J.~Ruvalds, Phys. Rev. B
{\bf 60}, 12432 (1999).

\bibitem{pan'01} V.~M.~Pan, V.~S.~Flis, V.~A.~Komashko, O.~P.~Karasevska,
V.~L.~Svetshnikov, M.~Lorenz, A.~N.~Ivanyuta, G.~A.~Melkov, E.~A.~Pashitskii, and H.~W.~Zandbergen, IEEE Trans. Appl. Supercond. {\bf 11}, 3960 (2001).

\bibitem{pan'03} V.~M.~Pan, C.~G.~Tretiatchenko, V.~S.~Flis, V.~A.~Komashko,
E.~A.~Pashitskii, A.~N.~Ivanyuta, G.~A.~Melkov, W.~H.~Zandbergen, and
V.~L.~Svetchnikov, J. Supercond. {\bf 16}, 889 (2003).

\bibitem{Streif'91} S.~K.~Streiffer, B.~M.~Lairson, C.~B.~Eom, B.~M.~Clemens,
J.~C.~Bravman, and T.~H.~Geballe, Phys. Rev. B {\bf 43}, 13007 (1991).

\bibitem{chish'92} S.~J.~Pennycook, M.~F.~Chisholm, D.~E.~Jesson, R.~Feenstra,
S.~Zhu, X.~Y.~Cheng, and D.~J.~Lowndes, Physica C {\bf 202}, 1 (1992).

\bibitem{Merkle'91} Y.~Gao, K.~L.~Merkle, G.~Bai, H.~L.~Chang, and D.~J.~Lam,
Physica C {\bf 174}, 1 (1991).

\bibitem{pan'97} V.~L.~Svetchnikov, V.~M.~Pan, Ch.~Traeholt, and W.~H.~Zandbergen, IEEE Trans. Appl. Supercond. {\bf 7}, 1396 (1997).

\bibitem{Dam'99} B.~Dam, J.~M.~Huijbregtse, F.~C.~Claassen, R.~C.~F.~van der Geest, G.~Doornbos, J.~H. Rector, A.~M.~Testa, S.~Freisem, J.~C.~Martinez, B.~Stauble-Pumpin, and R.~Griessen, Nature (London) {\bf 399}, 439 (1999).

\bibitem{Dam'00} J.~M.~Huijbregtse, B.~Dam, R.~C.~F.~van der Geest, F.~C.~Klaassen, R.~Elberse, J.~H.~Rector, and R.~Griessen, Phys. Rev. B {\bf 62}, 1338 (2000).

\bibitem{Dam'02} B.~Dam, J.~M.~Huijbregtse and J.~H.~Rector, Phys. Rev. B
{\bf 65}, 064528 (2002).

\bibitem{Cas'99} A.~Casaca, G.~Bonfait, C.~Dubourdieu, F.~Weiss, and J.~P.~Senateur, Phys.Rev. B {\bf 59}, 1538 (1999).

\bibitem{Dur'01} S.~Berger, D.-G.~Crete, J.-P.~Contour, K.~Bouzehouane,
J.-L. Maurice, and O. Durand, Phys. Rev. B {\bf 63} 144506 (2001).

\bibitem{Cao'02} L.~X.~Cao, T.~L.~Lee, F.~Renner, Y.~X.~Su, R.~L.~Johnson, and J.~Zegenhagen, Phys. Rev. B {\bf 65}, 113402 (2002).

\bibitem{Rob'03} J.~J.~Robles, A.~Bartasyte, H.~P.~Ng, A.~Abrutis, and F.~Weiss, Physica C {\bf 400}, 36 (2003). 

\bibitem{Tinkham'82} G.~E.~Blonder, M.~Tinkham, and T.~M.~Klapwijk, Phys. Rev. B {\bf 25}, 4515 (1982).

\bibitem{Tinkham'96} M.~Tinkham, {\it Inroduction to Superconductivity}
(McGraw-Hill, Inc., New York, 1996).

\bibitem{Scalapino'95} D.~J.~Scalapino, Phys. Rep. {\bf 250}, 329 (1995).

\bibitem{Kirtley'00} C.~G.~Tsuei and J.~R.~Kirtley, Rev. Mod. Phys. {\bf 72},
969 (2000).

\bibitem{Yeh'98} J.~Y.~T.~Wei, N.~C.~Yeh, D.~F.~Garrigus, and M.~Strasik, Phys.
Rev. Lett. {\bf 81}, 2542 (1998).

\bibitem{Abr} A.~A.~Abrikosov, {\it Fundamentals of the Theory of Metals} (Elseiver Science, Amsterdam, 1988).

\bibitem{Sch'98} M.~C.~Schabel, C.-H.~Park, A.~Matsuura, Z.-X.~Shen, D.~A.~Bonn, R.~Liang, and W.~N.~Hardy, Phys. Rev B {\bf 57}, 6090 (1998).

\bibitem{Jar'00} T.~Jarlborg and G.~Santi, Physica C {\bf 329}, 243 (2000).
   
\bibitem{galperin'86} A.~G.~Aronov, Yu.~M.~Galperin, V.~L.~Gurevich, and
V.~I.~Kozub, in {\it Nonequilibrium Superconductivity}, edited by
D.~N.~Langenberg and A.~I.~Larkin (Elsevier Science Publishers B.V., Amsterdam,
1986), p. 325; Adv. Phys. {\bf 30}, 539 (1981).

\bibitem{Blatter'99} G.~Blatter, V.~B.~Geshkenbein, and N.~B.~Kopnin, Phys. Rev. B 
{\bf 59}, 14663 (1999).

\bibitem{Kopnin'01} N.~B.~Kopnin, J. Low Temp. Phys. {\bf 124}, 209 (2001).

\bibitem{Ziman'72} J.~M.~Ziman, {\it Principles of the Theory of Solids} (University Press, Cambridge, 1972).

\end{thebibliography}
\end{document}